\documentclass[
prl,
superscriptaddress,amsmath,amssymb,
twocolumn,
]{revtex4-1}
    
\usepackage{graphicx}
\usepackage{dcolumn}
\usepackage{bm}
\usepackage{tikz}
\usepackage[normalem]{ulem}
\usepackage{blindtext}
\usepackage{tabularx}
\usepackage{amsmath}
\usepackage{xr}
\usepackage{mathtools}
\usepackage{changes}
\usetikzlibrary{calc}
\usetikzlibrary{fit}
\usetikzlibrary{decorations.pathmorphing}
\usetikzlibrary{arrows,shapes,positioning}
\usetikzlibrary{decorations.markings}
\tikzstyle arrowstyle=[scale=1]
\tikzstyle directed=[postaction={decorate,decoration={markings,
    mark=at position .65 with {\arrow[arrowstyle]{stealth}}}}]
\tikzstyle reverse directed=[postaction={decorate,decoration={markings, mark=at position .65 with {\arrowreversed[arrowstyle]{stealth};}}}]
\makeatletter
\newcommand*{\addFileDependency}[1]{
  \typeout{(#1)}
  \@addtofilelist{#1}
  \IfFileExists{#1}{}{\typeout{No file #1.}}
}
\makeatother
\newcommand*{\myexternaldocument}[1]{%
    \externaldocument{#1}%
    \addFileDependency{#1.tex}%
    \addFileDependency{#1.aux}%
}
\myexternaldocument{Supplementary}

\usepackage{color}

\newcommand{\rucl}{$\alpha$-RuCl$_3$\xspace}

\newcommand{\jeff}{$j_{\text{eff}}$\xspace}
\newcommand{\crc}{Cl--Ru--Cl trilayers\xspace}

\begin{document}

\title{Pressure-tuning of \rucl towards the ideal Kitaev-limit}  

\author{Q.~Stahl}
\affiliation{Institut f\"ur Festk\"orper- und Materialphysik, Technische Universit\"at Dresden, 01062 Dresden, Germany}

\author{T.~Ritschel}
\affiliation{Institut f\"ur Festk\"orper- und Materialphysik, Technische Universit\"at Dresden, 01062 Dresden, Germany}

\author{G.~Garbarino}
\affiliation{European Synchrotron Radiation Facility, 38042 Grenoble, France}

\author{F. Cova}
\affiliation{European Synchrotron Radiation Facility, 38042 Grenoble, France}


\author{A. Isaeva}
\affiliation{Fakult\"at f\"ur Chemie und Lebensmittelchemie, Technische Universit\"at Dresden, 01062 Dresden, Germany}
\affiliation{Van der Waals - Zeeman Institute, Institute of Physics, University of Amsterdam, 1098 XH Amsterdam, The Netherlands}

\author{T. Doert}
\affiliation{Fakult\"at f\"ur Chemie und Lebensmittelchemie, Technische Universit\"at Dresden, 01062 Dresden, Germany}

\author{J.~Geck}
\affiliation{Institut f\"ur Festk\"orper- und Materialphysik, Technische Universit\"at Dresden, 01062 Dresden, Germany}
\affiliation{W\"urzburg-Dresden Cluster of Excellence ct.qmat, Technische Universit\"at Dresden, 01062 Dresden, Germany}

\date{\today}

\begin{abstract}

We report the discovery of an intriguing pressure-driven phase transformation in the layered Kitaev-material \rucl.  
By analyzing both the Bragg scattering as well as the diffuse scattering of high-quality single crystals, 
we reveal a collective reorganization of the layer stacking throughout the crystal. Importantly, this transformation also effects the structure of the RuCl$_3$ honeycomb layers, which acquire a  high trigonal symmetry with a single Ru--Ru distance of 3.41\,\AA\/ and a single Ru--Cl--Ru bond angle of 92.8$^\circ$. Hydrostatic pressure therefore allows to tune the structure of \rucl much closer to the ideal Kitaev-limit. The high-symmetry phase can also be stabilized by biaxial stress, which can explain conflicting results reported earlier and, more importantly, makes the high-symmetry phase accessible to a variety of experiments.

\end{abstract}

\maketitle

\section*{Introduction}

Quantum spin liquids (QSLs) are fascinating states of matter in which competing interactions between magnetic moments prevent static magnetic order down to even zero temperature\,\cite{Broholm:2020p, Takagi2019, Balents2010}. Instead a massive quantum entanglement of spins dominates. QSLs can therefore not be identified by broken symmetries, nor do they correspond to a trivial disordered spin system. Instead, QSLs exhibit non-trivial topological properties, many-body quantum entanglement and emergent fractionalized quasiparticles.

A specific and very famous example is the so-called Kitaev-QSL, where the constituent spins can fractionalize into mobile Majorana fermions coupled to conserved $\mathbb{Z}_2$-fluxes\,\cite{Kitaev:2006aa}. Applying a magnetic field even results in a non-abelian QSL, which may in fact be a key ingredient for topological quantum computing\,\cite{Nayak:2008x} and certainly is one reason for the large interest in these systems. As far as specific materials are concerned, a number of candidates could already be identified, including Na$_2$IrO$_3$, different polytypes of Li$_2$IrO$_3$, H$_3$LiIr$_2$O$_6$ and \rucl\,\cite{Takagi2019}. Among these, as a matter of fact, \rucl turned out to be particularly promising, because earlier indications of a field-induced QSL\,\cite{Baek:2017a,Wolter:2017a} appear be supported by signatures of a quantized thermal Hall conductance\,\cite{Yokoi:2021n}.

\begin{figure*}
\centering
\includegraphics[width=\textwidth]{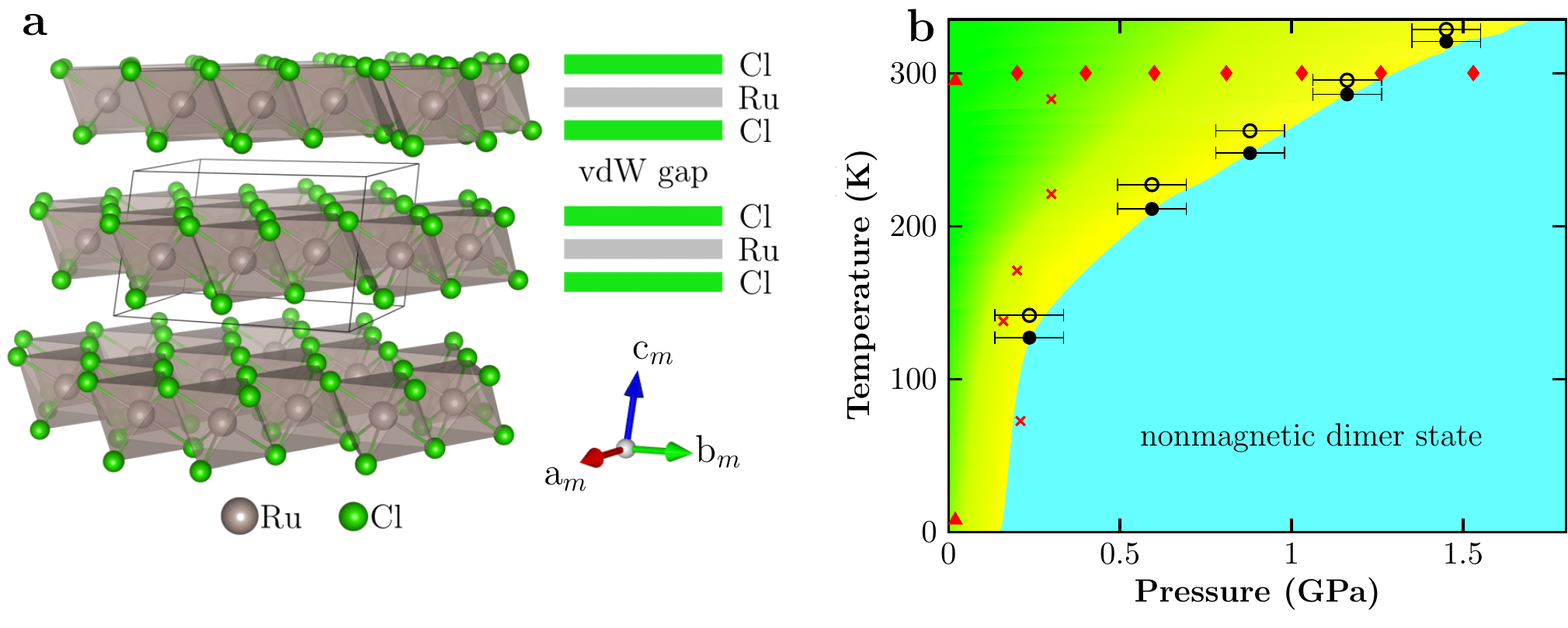}
\caption{\textbf{Structure and layer stacking in \rucl as functions of temperature and pressure.} (a) Illustration of the three-dimensional monoclinic $C2/m$ structure of \rucl at ambient conditions. (b) Pressure-temperature phase diagram of \rucl. The red triangles and diamonds indicate points where full single-crystal diffraction data sets have been recorded at ambient  and at increased pressure, respectively. The red crosses indicate points where overview scans have been collected (further explanations in the text). The $C2/m$ symmetry with cubic close packed Cl is preserved in stress-free samples down to 3\,K, as illustrated by the green region. Yellow indicates the region where the phase with hexagonal close packed Cl is stable. With further increasing pressure, a nonmagnetic dimer state is reached. The solid and open black circles, taken from Ref.\,\onlinecite{Bastien2018}, mark the reduction of the magnetic susceptibility upon cooling and warming at constant $p$, respectively.}
\label{fig:intro}
\end{figure*}

\rucl is a spin-orbit assisted Mott-insulator with honeycomb layers formed by edge-sharing RuCl$_3$-octahedra \,\cite{Plumb2014}, as illustrated in Fig.\,\ref{fig:intro}\,(a). The low-energy magnetism of the honeycomb layers can be described in terms of \mbox{\jeff\,=\,1/2} pseudo-spins of Ru $4d^5$\cite{Yadav:2016b}. However, \rucl is not a pure Kitaev-system. Besides the Kitaev-interactions, there are also other magnetic couplings beyond the ideal Kitaev-model, such  as the Heisenberg exchange and the off-diagonal symmetric exchange\,\cite{Kaib:2021aa,Rau:2014aa}. Due to these additional interactions, \rucl does not display a Kitaev-QSL at ambient pressure and low temperature, but instead antiferromagnetic zigzag-order below $T_N=7$\,K\,\cite{Cao:2016a}.  
It has been realized early on that the magnetic order of \rucl is affected by the stacking order of the RuCl$_3$ layers:
A single and well-defined $T_N=7$\,K was found to require a well-defined layer-stacking, while disordered stacking broadens the magnetic transition significantly and can even introduce a second magnetic transition at 14\,K\,\cite{Cao:2016}.

In general, layered transition metal trihalides are well-known for their polytypism\,\cite{Ruck:2000}, meaning in particular that different stackings of the strongly covalent layers can occur. The polytypism facilitates modification of the stacking via external parameters, which in turn can provide a handle to manipulate magnetic properties\,\cite{song2019,li2019,chen2019}. This is a strong motivation for exploring the relevance of polytypism to magnetism in the Kitaev-material \rucl.

One way to trigger different polytypes of \rucl via an external parameter is to apply hydrostatic pressure $p$. Previous studies of \rucl already established that $p$ in fact stabilizes a broken-symmetry state with ordered Ru--Ru dimers, i.e., a valence bond crystal\,\cite{Bastien:2018a,biesner2018}, see Fig.\,\ref{fig:intro}\,(b).
Here we study the transition from the ambient pressure monoclinic phase into this dimerized phase at room temperature in more detail. By means of x-ray diffraction (XRD) we discover that applying pressure at constant temperature $T$ stabilizes a rhombohedral phase of \rucl with short-range stacking order, located in between the ambient and the dimerized phase in the $pT$ phase diagram. Importantly, the RuCl$_3$ layers in this rhombohedral phase acquire a high-symmetry configuration, bringing their geometry much closer to the ideal Kitaev-limit where all the Ru--Cl--Ru bond angles are 90$^\circ$ and all the Ru--Ru as well Ru--Cl distances are identical.

\section{Experimental details}

\rucl single crystals were grown from phase-pure commercial \rucl powder via a high-temperature vapor transport technique and carefully characterized as described previously\,\cite{Bastien2018}. We confirmed the monoclinic $C$2/$m$ structure at ambient conditions for a disorder-free crystal (crystal 1) by means of single crystal XRD (for details see Appendix). The obtained crystallographic parameters are fully consistent with previously published results \cite{Cao:2016a}. 
 
High-pressure XRD studies were carried out at beamlines ID15B and ID27 of the European Synchrotron Radiation Facility (ESRF) in Grenoble. To provide nearly hydrostatic pressure conditions the membrane-driven diamond anvil cell (DAC) was loaded with helium as pressure transmitting medium. The pressure inside the DAC was monitored {\it in-situ} using the R$_{1,2}$ fluorescence of Cr-centers in ruby spheres placed next to the sample. The high-pressure XRD was done in transmission geometry with the DAC mounted on a single-axis goniometer with the rotation axis ($\omega$)  perpendicular to the scattering plane. The diffracted radiation was recorded with a MAR555 flat panel detector at ID15B and a MAR-CCD detector at ID27 installed perpendicular to the primary beam. 

Single crystal XRD data at room temperature were collected at ID15B, using a monochromatic radiation of 30\,keV (\mbox{$\lambda$=0.4113\,\AA}) and a spot size of 10x10\,$\mu$m$^2$ [red diamonds in Fig.\,\ref{fig:intro}\,(b)]. Diffraction data were recorded in steps of approximately 0.2\,GPa up to 2\,GPa. Each data set contains 120 frames with 0.5$^\circ$ scan width and an exposure time of 1\,s per frame over a sample rotation of 60$^\circ$ (-30$^\circ$ $\leq$ $\omega$ $\leq$ 30$^\circ$). 
Using a very similar experimental setup at beamline ID27, the structural pressure-temperature phase diagram of \rucl was mapped out further at lower temperatures as well, using a continuous He-flow cryostat [red crosses in Fig.\,\ref{fig:intro}\,(b)]. 
In this experiment, the single crystalline sample was exposed to a monochromatic 3x3\,$\mu$m$^2$ x-ray beam with a photon energy of 33\,keV (\mbox{$\lambda$=0.3738\,\AA}), while continuously recording the diffracted intensity on the detector during an $\omega$-movement of 60$^\circ$ (-30$^\circ$ $\leq$ $\omega$ $\leq$ 30$^\circ$).

\begin{figure*}
 \includegraphics[width=1.0\textwidth]{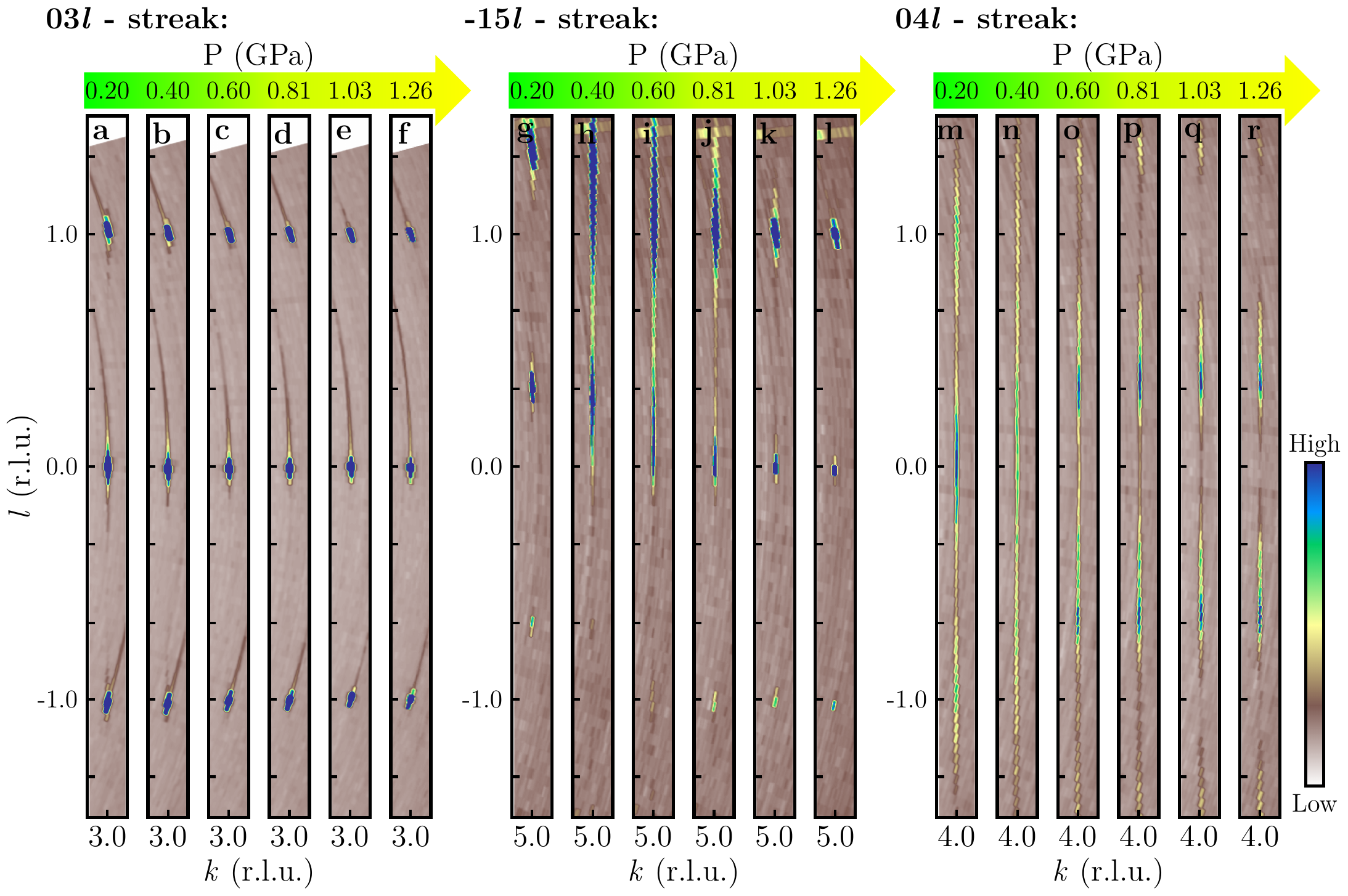}
 \caption{\textbf{Evolution of the XRD pattern during the pressure-driven phase transition.} Reciprocal space maps parallel to the 03$l$-plane (a)-(f), the -15$l$-plane (g)-(l) and the 04$l$-plane (m)-(r) (integration thickness perpendicular to the plane: $\Delta h$ = 0.04) in the pressure range 0.2\,GPa to 1.26\,GPa at 300\,K. The indexation corresponds to the hexagonal setting (cf. Fig.\,\ref{fig:average}). The diffuse stripes along the $l$-direction observable in (h)-(j) and (m)-(r) are a hallmark of stacking faults. The intensity shift along the $l$-direction in (g)-(l) is representative for all reflections with $h-k=3n$ ($n$ integer and $h$, $k \neq 3n$) and reveals the rearrangement of the Cl--Ru--Cl sandwich layers from a cubic to a hexagonal closed chlorine packing. While, the diffuse stripes along the $l$-direction fulfilling the condition $h-k=3n\pm1$ (with $n$ integer) as depicted in (m)-(r)  signal the disordered stacking of the Ru honeycomb nets over the entire pressure range.}
\label{fig:pdep}
 \end{figure*} 

\section{Results}

Fully consistent with previous reports, our diffraction patterns taken at ambient conditions can be indexed by a monoclinic unit cell (space group $C2/m$) with lattice parameters a$_m$=5.9875(6)\,\AA, b$_m$=10.3529(3)\,\AA, c$_m$=6.0456(6)\,\AA~and $\beta$=108.777(9), see Fig.\,\ref{fig:intro}\,(a). 
However, the crystal studied as a function of $p$ at room temperature (crystal 2) showed, apart from sharp Bragg reflections, a set of one-dimensional diffuse scattering rods at zero pressure, cf. Fig.~\ref{fig:pdep}. These rods are oriented parallel to c$^*_m$, thus revealing a disordered stacking of the \crc along the c$_m$ direction, which can be attributed to stacking faults, characterized by b$_m$/3 shifts
between adjacent layers\,\cite{Johnson:2015}. 

\begin{figure}
 \includegraphics[width=\columnwidth]{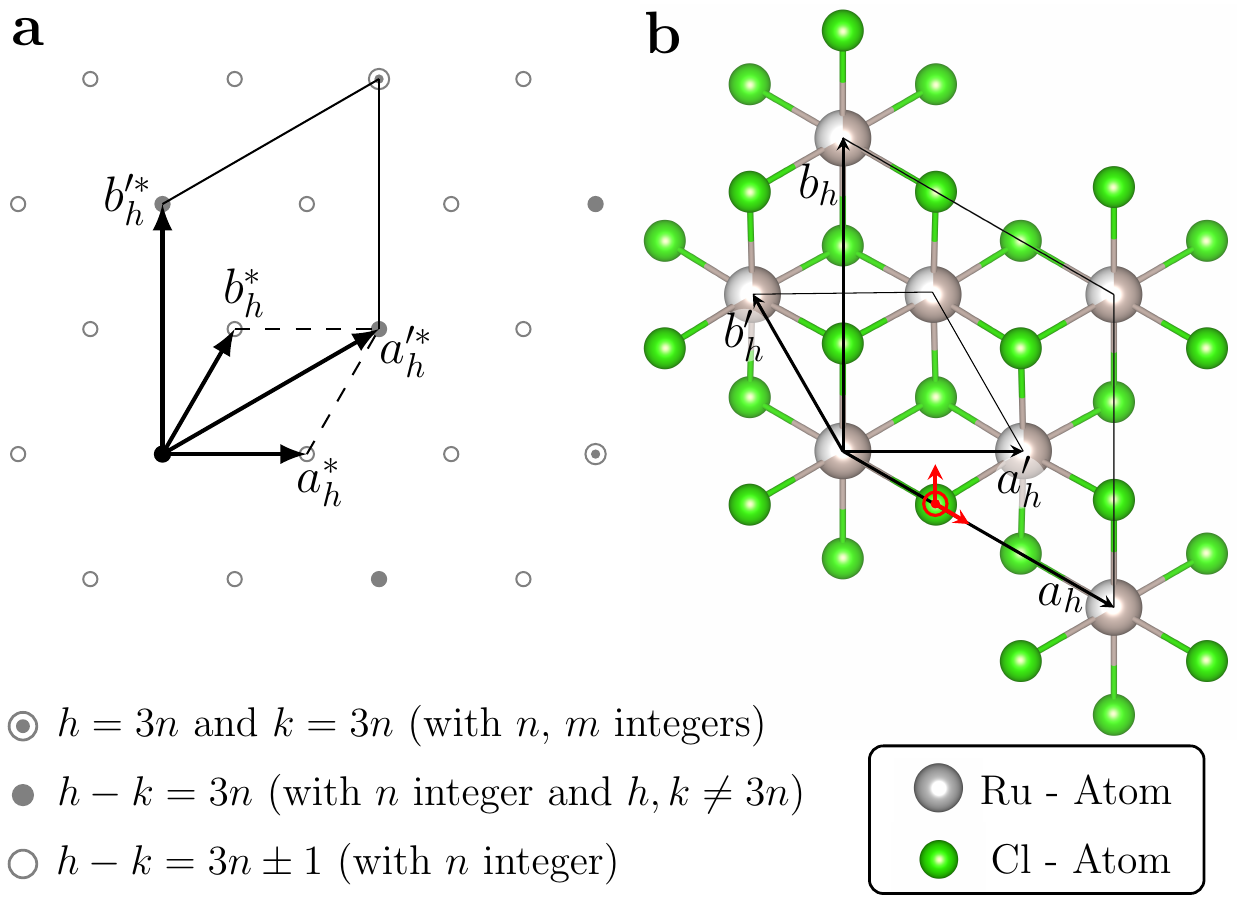}
 \caption{\textbf{Schematic illustration of the $hk$0-layer and the relation between the real and the average layer structure of \rucl at 1.26GPa.}
 (a) The location of diffuse rods and sharp reflections as a function of pressure leads to three distinguishable families of reflections, drawn in by miscellaneous symbols. The schematic diffraction pattern showing the reciprocal lattice vectors for both the average ($a'^{*}_{h}$,$b'^{*}_{h}$) and the real ($a^{*}_{h}$, $b^{*}_{h}$) layer structure.
(b) The average layer structure deduced from the sharp Bragg reflections is marked by the in-plane basis vectors \textbf{a’$_h$} and \textbf{b’$_h$}. The average structure consists of chlorine atoms in a hexagonal-close-packing arrangement in which the Ru atoms occupy all octahedral voids within a layer statistically with a site occupation factor of 2/3. The Cl-displacements $\delta x_{Cl}$ and $\delta y_{Cl}$ from the average position in the real layer structure are indicated by red arrows parallel to a$_h$  and b$_h$, respectively.
}
\label{fig:average}
\end{figure}

To describe such stacking faults and the polytypism in \rucl, it is convenient to use a hexagonal set of basis vectors a$_h$, b$_h$ and c$_h$, with a$_h$ and b$_h$ parallel to the Cl--Ru--Cl layer and the c$_h$-axis normal to it (cf. Fig.\,\ref{fig:average}). In doing so, we neglect that the hexagonal symmetry is broken by a small monoclinic distortion in some regions of the phase diagram. However, this approximation enables us to draw upon the theoretical framework for XRD of transition metal trihalides with stacking faults developed by Ulrich Müller and Elke Conradi\,\cite{Mueller1986}.

\subsection{Rearrangement of \crc}

In Fig.~\ref{fig:pdep} we show the evolution of the x-ray intensity distribution in reciprocal space during the pressure induced rearrangement of the \crc at room temperature. Here and throughout the following, the indexing of the reflections refers to the approximate hexagonal cell introduced in the previous paragraph. In order to reveal changes in the layer stacking along c$_h$, the intensities within a slice with $-0.02 \leq \Delta h \leq 0.02$ were projected onto the $kl$-plane. 

The reflections in these reciprocal space maps can be divided into three families, depending on $h$ and $k$: 

\textbf{Family 1:} $h=3n$ and $k=3m$ (with $n$, $m$ integers). This family is represented by the 03$l$-streak shown in Figs.~\ref{fig:pdep}~(a)-(f), where it can be observed that these reflections appear at integer $l$ positions and stay sharp along the $l$-direction over the entire pressure range shown. 

\textbf{Family 2:} $h-k=3n$ with $n$ integer and $h, k \neq 3n$. These peaks are represented by the  -15$l$-streak in Figs.~\ref{fig:pdep}~(g)--(l).
At 0.2\,GPa these reflections are sharp and centered at $l=1/3+n$. According to Ref.\,\cite{Mueller1986}, the position of this family of reflections provides information about the Cl packing. Specifically, reflections located at $l=1/3+n$ indicate a cubic close packing of Cl, which indeed corresponds precisely to the monoclinic $C2/m$ crystal structure of \rucl at ambient conditions. 
However, above 0.2\,GPa, a very pronounced broadening of the peaks is observable and intensity starts to spread along the $l$-direction, as can be seen in Figs.~\ref{fig:pdep}~(h)--(j). This signals the loss of long-range order of the Cl-sites in the out-of-plane direction. Note that the intensity maximum shifts towards integer $l$ values between 0.40\,GPa and 0.81\,GPa. Surprisingly, upon increasing $p$ further, the peaks again become sharper and sharper, until they are again resolution limited at 1.26\,GPa. 
The new long-range ordered state at this pressure is characterized by the reflection condition $l=n$ (with $n$ integer) and peak widths comparable with the width of the 1st family. This reveals a rearrangement of the \crc, resulting in a transition from a cubic to a hexagonal chlorine close packing\,\cite{Mueller1986}. Interestingly, this $p$-driven transition between the two long-ranged ordered phases requires collective sliding of the \crc by about 2\,\AA.

\begin{figure}[t!]
 \includegraphics[width=.95\columnwidth]{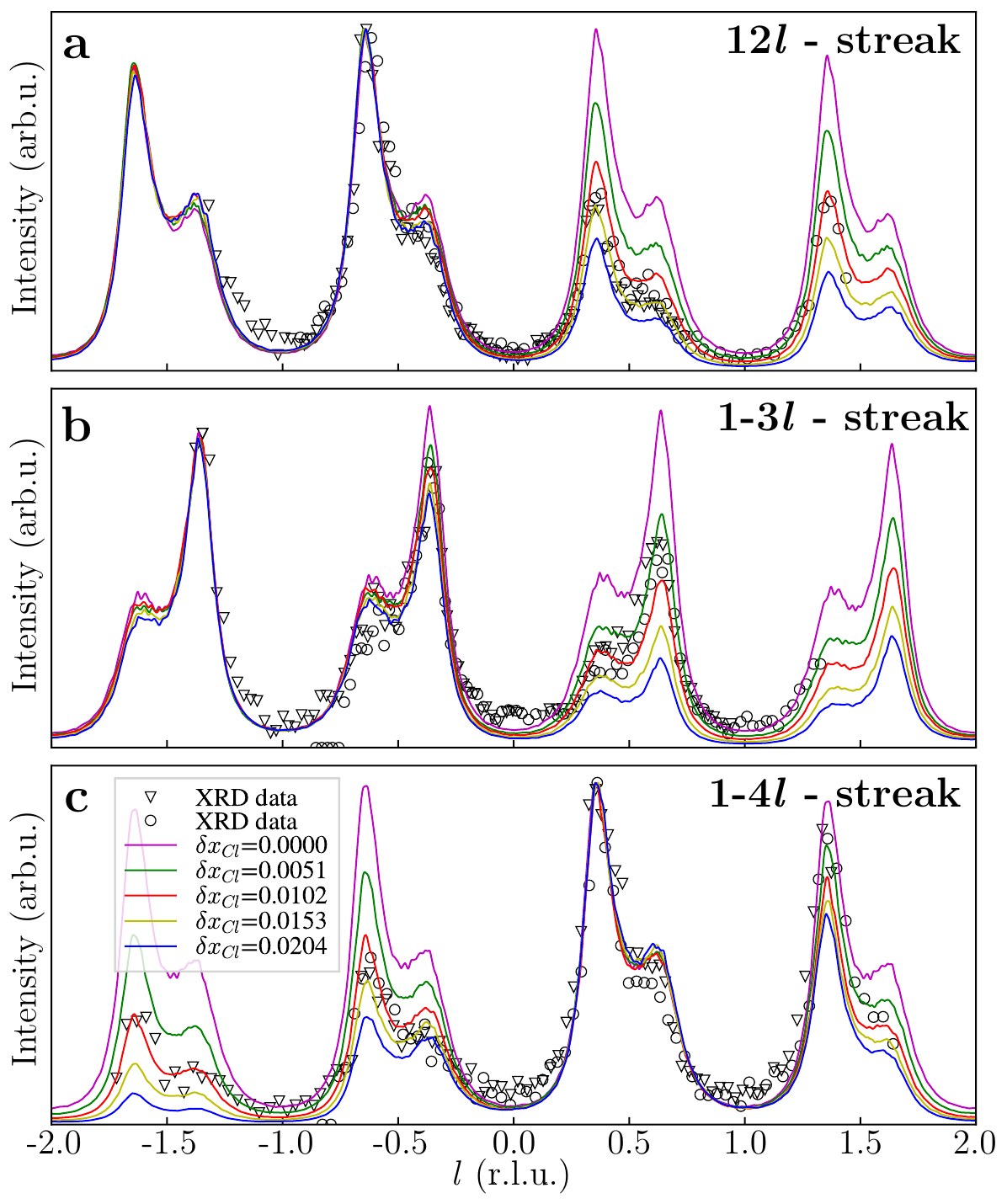}
 \caption{\textbf{Comparison of calculated intensity profiles with XRD data taken at 1.26 GPa and evolution.} Diffuse intensity profiles along the 12$l$ (a), 1-3$l$ (b) and 1-4$l$ (c) streaks (black circles) and the inverted symmetry-equivalent profiles (black triangles). The intensity profiles were corrected for background intensity as well Lorentz and polarization corrections were applied (see Appendix). 
 The $Cl$ atom is displaced in 3\,pm steps (equals $\delta x_{Cl} = 0.0051$ in fractional coordinates) parallel to $a_h$ from the average position $x_{Cl} = 1/3$ deduced from the sharp reflections. The Miller index $l$ refer to the hexagonal basis vector c$_h$. 
}
\label{fig:hexagonal}
 \end{figure}

\textbf{Family 3:} $h-k=3n\pm1$ with  $n$ integer. This family is represented in Figs.~\ref{fig:pdep}~(m)--(r), which shows the intensity distribution along the 04$l$-streak. The intensity of these reflections is very broad and diffuse along $l$ already at 0.2\,GPa, but a maximum at $l=n$ is still recognizable. Referring again to Ref.\,\,\cite{Mueller1986}, this implies that the stacking of the Ru honeycomb layers along $c_h$ is disordered already at 0.2\,GPa, which, in fact, is also the case at ambient pressure for this sample (crystal 2). Notwithstanding, the maxima at $l=n$ at ambient pressure show that this sample exhibits the expected $C2/m$-structure at ambient pressure, although with stacking faults. Note that, even though the stacking of the Ru honeycomb layers is always disordered in this crystal, the stacking of the Cl-layers at ambient pressure and at $p=1.26$\,GPa is fully ordered in all spatial directions. This is due to the fact that for fixed Cl-positions the Ru-layers, which are sandwiched between the Cl-layers, can still assume different positions. 
Upon increasing pressure, the intensity distribution initially becomes completely smeared out along $l$, but then, with further increasing pressure, it accumulates in broad maxima at $l=n+1/3$ and $l=n+2/3$.
Further representative intensity distributions for the $3^{rd}$ family of reflections at 1.26\,GPa are given in Fig.~\ref{fig:hexagonal}~(a)--(c), where the diffuse maxima for different $h-k=3n\pm1$ are shown.

Taken together, the data presented in Fig.~\ref{fig:pdep} thus uncovers a pressure-induced rearrangement of the \crc, which results in an ordered hexagonal close packing of chlorine atoms, but lacks a long-range ordered stacking of the Ru honeycomb layers along $c_h$.

\subsection{Structure of the \crc}

In order to determine the structure of the \crc, we analyze both the sharp Bragg reflections and diffuse scattering. In Fig.~\ref{fig:average}\,(a) the locations of the different reflection families are illustrated. The first step is to determine the averaged structure of a single Cl--Ru--Cl layer at room temperature and $p=1.26$\,GPa from the sharp Bragg peaks alone (family 1 and 2). This intensity distribution can be modeled in terms of the averaged RuCl$_3$-layer shown in Fig.\,\ref{fig:average}\,(b), which corresponds to a projection of all atomic sites along c$_h$-direction onto a single layer. Note that the Ru-sites in this average layer do not form a honeycomb net and that all Ru-sites posses a site occupation factor of 2/3. 

These layers are then stacked on top of each other along $c'_h$ to form an averaged three dimensional structure.
Analysing the positions of the sharp Bragg peaks only, we find that the three dimensional averaged structure is described by a trigonal cell with lattice parameters $a'_h$=$b'_h$ = 3.4080(4)\,\AA~and $c'_h$=5.562(10)\,\AA.
%
While no systematic extinction condition was found, the analysis of the intensities is consistent with the trigonal Laue class $\bar3$\textit{m}1. The averaged structure was then solved and refined in the space group \textit{P}$\bar3$\textit{m}1 as described in detail in the Appendix. On account of the small atomic displacements and possible strong parameter correlations in fact no split positions but merely averaged atomic positions are obtained by our crystallographic refinements.

From the atomic positions determined in this way, the real structure of the honeycomb \crc can then be reconstructed straightforwardly, assuming a fixed stoichiometry. This yields
Ru (1a) at $(2/3,1/3,1/2+\delta z_{Ru})$ and Cl (1a) at $(1/3+\delta x_{Cl},\delta y_{Cl},0.740)$. 

The layer-group symmetry of the above model is \textit{p}$\bar3$1\textit{m} for $\delta z_{Ru}= \delta y_{Cl}=0$. Let us consider possible symmetry reductions due to a finite  $\delta z_{Ru}$ or $\delta y_{Cl}$: 
$\delta z_{Ru}\neq0$ in combination with layer disorder discussed above, inevitably results in diffuse streaks along $l$ for \textit{all} combinations of the Miller Indices $h$ and $k$ at 1.26\,GPa, which is at variance with the results presented in Fig.\,\ref{fig:pdep} for family 1. From the absence of diffuse streaks for all reflections with $h=3n$ and $k=3m$ we therefore conclude $\delta z_{Ru}=0$. As described previously\,\cite{Ruck1995}, a finite $\delta y_{Cl}>0$ would break the symmetry around the Ru-site, which contradicts the condition $\delta z_{Ru}=0$. The latter therefore also implies that $\delta y_{Cl}$ must vanish as well. 
As a result, we obtain a trigonal cell with lattice parameters $a_h$=$b_h$= 5.9028(4)\,\AA~and $c_h$=5.562(10)\,\AA. The asymmetric unit of this structure contains one Ru-site at $(2/3,1/3,1/2)$ and one Cl-site at $(1/3+\delta x_{Cl},0,0.740)$.

\begin{table*}
  \begin{ruledtabular}
    \begin{tabular}{cccccc}
	\multicolumn{2}{c}{} & \multicolumn{2}{c}{0\,GPa -  \textit{C}2/\textit{m}} & \multicolumn{2}{c}{1.26\,GPa - \textit{p}$\bar3$1\textit{m}}\\
	\multicolumn{2}{c}{} & Experiment & DFT & Experiment & DFT\\
	\hline
	Ru &  x &  0 & 0 & 2/3 & 2/3\\
	     &  y &  0.16651(2) & 0.16635 & 1/3 & 1/3\\
	     &  z &  1/2 & 1/2 & 1/2 & 1/2\\
	Cl$_{1}$ &  x &  0.22680(13) & 0.22692 & 0.3435(17) & 0.34291\\
	     &  y &  0 & 0 & 0 & 0.00022\\
	     &  z &  0.73488(12) & 0.73488 & 0.7400(20) & 0.74045\\
	Cl$_{2}$ &  x &  0.25058(10) & 0.25059 &  & \\
	     &  y &  0.17411(4) & 0.17407 &   & \\
	     &  z &  0.26761(9) & 0.26769 &   & \\
	\hline
	\multicolumn{2}{c}{Ru-Ru\,(\AA)} & 3.4477(5) / 3.4570 (4) & 3.4444 / 3.4587 & 3.4080(3) & 3.4080 \\
	\multicolumn{2}{c}{Ru-Cl-Ru\,($^\circ$)} & 93.62(3) / 94.04(3) & 93.55 / 94.08 & 92.8(4) & 92.62 \\
    \end{tabular}
    \caption{\textbf{Structural parameters as determined from the single crystal data and structural optimization calculations as a function of pressure at 300\,K.}
    At ambient pressure, the lattice parameter are a$_m$=5.9875(6)\,\AA, b$_m$=10.3529(3)\,\AA, c$_m$=6.0456(6)\,\AA~and $\beta$=108.777(9) and the structure is described by the space group $C$2$/m$.
    At 1.26\,GPa the lattice parameter are a$_h$=b$_h$=5.9028(4)\,\AA and c$_h$=5.562(10)\,\AA~and the structure of a single Cl-Ru-Cl layer is described by the trigonal layer group \textit{p}$\bar3$1\textit{m}. We note that the fractional coordinate we obtain from structural optimization calculations for the chlorine atom very slightly deviates from the \textit{p}$\bar3$1\textit{m} symmetry. Illustrations of the ambient and high pressure structural models are shown in Fig.~\ref{fig:StrucTrans}\,(a), (b) and Fig.~\ref{fig:StrucTrans}(c), (d) respectively.
    }
    \label{tab:par_EXP_DFT}
  \end{ruledtabular}
\end{table*}

\subsection{Analysis of diffuse scattering and structure model}

For the determination $\delta x_{Cl}$ 
at $p=1.26$\,GPa, a quantitative analysis of the diffuse scattering along $l$ is needed. To this end, the experimental data is compared to the diffuse scattering of disordered model structures,
which were constructed starting from the \crc determined in the previous section. For fixed $\delta x_{Cl}$, stacks of 1000 \crc along the $c_h$-direction were generated in a first order Markov process in a way as to form hexagonal closed packed Cl-layers. Further details about the construction can be found in the Appendix. The displacement $\delta x_{Cl}$ was then determined by fitting the  model to the measured intensity profiles.

In Fig.~\ref{fig:hexagonal} (a)--(c) we show representative intensity profiles of reflections belonging to family 3 at 1.26\,GPa. As can be observed very nicely in this figure, the calculated diffuse intensity profiles depend very sensitively on $\delta x_{Cl}$, which enables its precise determination via comparison to experiment.
We find the best overall agreement between model and experiment for 
a Cl-displacement $\delta x_{Cl} = 0.0102$ [solid red lines in Fig.~\ref{fig:hexagonal} (a)--(c)].

The structural parameters determined from XRD for both the monoclinic phase at ambient pressure and the high-symmetry phase at $p$ = 1.26\,GPa are summarized in Table~\ref{tab:par_EXP_DFT}. In order to further substantiate our experimental findings, we  performed a structural optimization within density functional theory and the generalized gradient approximation (GGA) of the exchange-correlation potential using the QUANTUM-ESPRESSO software package\,\cite{Perdew.Ernzerhof.GeneralizedGradientApproximation.PRL.1996,Giannozzi.Baroni.Advancedcapabilitiesmaterials.JoPCM.2017, Giannozzi.Wentzcovitch.QUANTUMESPRESSOmodular.JoPCM.2009}. The
Kohn-Sham orbitals were expanded in a plane-wave basis set with a kinetic-energy cutoff of 100 Ry. The Brillouin zone was sampled on a grid of $24\times24\times24$ $k$-points and integrated with the optimized tetrahedron method~\cite{Kawamura.Tsuneyuki.Improvedtetrahedronmethod.PRB.2014}.
For the structural optimization, the lattice constants and the space group were kept fixed (C2/m at ambient pressure and R$\bar 3$ at 1.26\,GPa), while the total energy and internal forces were optimized as a function of the allowed fractional coordinates.  The results of these calculations are also included in Table~\ref{tab:par_EXP_DFT} where they can be compared directly to the experimental values. We find excellent agreement between DFT and experiment, which confirms the experimentally determined structures and fractional coordinates.

\subsection{Discussion and conclusion}

\begin{figure} [b!]
\includegraphics[width=\columnwidth]{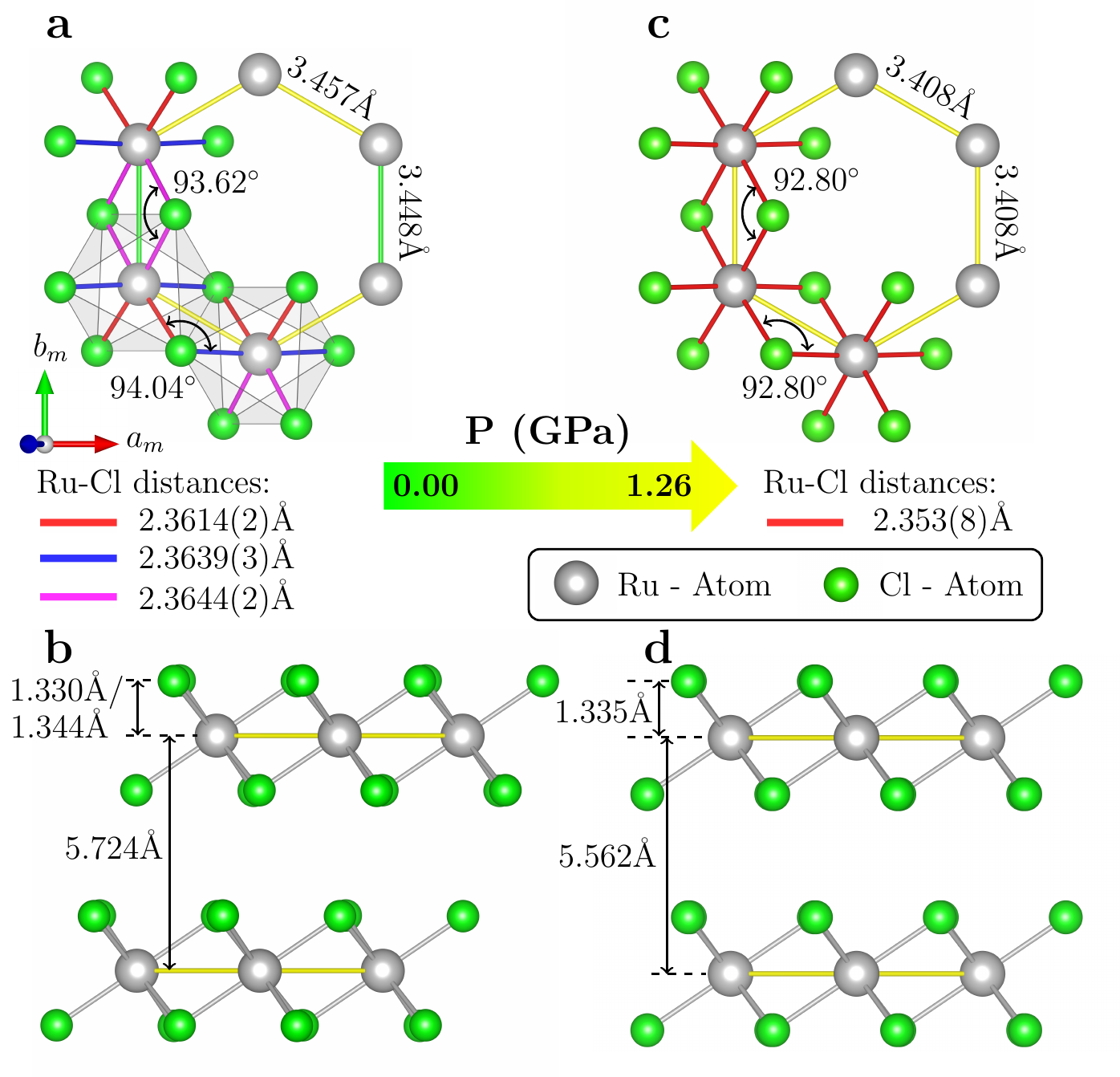}
\caption{\textbf{The crystal structure of \rucl determined by analyzing the Bragg scattering as well as the diffuse scattering of high-quality single crystals at 0.0\,GPa and 1.26\,GPa.} The local Cl-Ru-Cl layer geometry is depicted in (a),(c) and two successive honeycomb layers viewed normal to the stacking direction are shown in (b),(d) for the monoclinic C2/m structure at 0.0\,GPa and the high-symmetry structure with \textit{p}$\bar3$1\textit{m} layer symmetry at 1.26\,GPa, respectively. Equivalent nearest neighbor Ru-Ru links and Ru-Cl bond distances are encoded in the same colors in (a) and (b). For improved clarity only two nearest neighbor edge-sharing RuCl$_6$ octahedra are shown in (a).}
\label{fig:StrucTrans}
\end{figure}

The pressure-driven structural transformation discovered here is illustrated further in Fig.~\ref{fig:StrucTrans}. As reported earlier, the structure of \rucl at ambient conditions deviates from the  ideal honeycomb structure, as there are different Ru--Cl distances, different Ru--Ru distances and different Ru--Cl--Ru bond angles [Figs.\,\ref{fig:StrucTrans}\,(a),(b)]. Upon increasing the pressure to 1.26\,GPa, these differences disappear: there is only one single Ru--Cl distance, one single Ru--Ru distance and one single Ru--Cl--Ru bond angle, as shown in Figs.\,\ref{fig:StrucTrans}\,(c),(d). The \crc of the pressure-induced \textit{p}$\bar3$1\textit{m} phase form an undistorted Ru-honeycomb net with a Ru--Cl--Ru bond angle of 92.80$^\circ$ -- very close to the ideal value of 90$^\circ$.
Comparing Fig. 5 (b) and (d), it is obvious that the applied pressure mostly affects the van der Waals spacing, while the thickness of a single layer remains almost the same. Consequently, no further trigonal distortion of the RuCl$_6$ octahedron occurs, leaving the \mbox{\jeff\,=\,1/2} state stable. Application of hydrostatic pressure therefore drives the structure of  the \crc closer to the ideal geometry for realizing the Kitaev-model.

According to the data shown in Fig.\,\ref{fig:pdep}, the $p$-induced high-symmetry phase exhibits a well defined trigonal layer symmetry \textit{p}$\bar3$1\textit{m}, but also a significant amount of stacking disorder of the Ru-honeycomb layers [Figs.\,\ref{fig:pdep}\,(m)--(r)]. The overall structure does therefore not correspond to an ordered  $R\overline{3}$-structure, but rather consist of structural $R\overline{3}$-domains separated by stacking faults. We will therefore refer to this phase as high-symmetry phase with \textit{p}$\bar3$1\textit{m} layer symmetry. 

\begin{figure}
\includegraphics[width=\columnwidth]{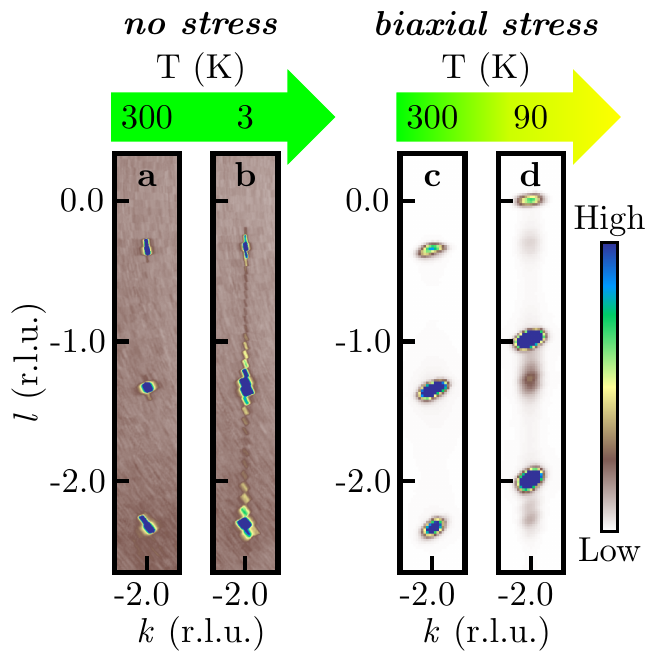}
\caption{
\textbf{Structural phase transition from monoclinic to rhombohedral symmetry driven by biaxial stress in bulk \rucl.} Reconstructed reciprocal space maps parallel to the 4-2$l$-plane (family 2) for the stress-free sample 1 (a),(b) and the sample 2 under biaxial stress are shown (c),(d). Peaks at $l=n-1/3$ and $l=n$ signal a cubic and a hexagonal closed chlorine packing, respectively.
}
\label{fig:strain}
\end{figure}

This high-symmetry polytype  has been determined at RT, which immediately raises the question as to whether it also exists at low-temperature. To address this issue we performed low-temperature measurements with two differently mounted samples: Sample\,1 was simply placed on a diamond anvil and \textit{not} glued down, in order to avoid external stresses caused by the different thermal contractions of sample and holder during cooling. Sample\,2 was instead firmly glued onto an Al-holder. At low $T$, the different thermal contractions of \rucl and Al do result in a biaxial stress parallel to the $ab$-plane of sample\,2.

As can nicely be observed in Fig.\,\ref{fig:strain}, the peak positions of the stress-free sample\,1 do not change upon cooling, indicating that this sample remains in the monoclinic $C2/m$ phase. For sample\,2, however, we do observe additional strong reflections at integer $l$ at low temperature, which signals an almost complete transition from the monoclinic to the high-symmetry phase with with \textit{p}$\bar3$1\textit{m} layer symmetry.
While our low-temperature data under biaxial stress does not allow for a detailed structure analysis, they are, however, fully consistent with the results published in Ref.\,\cite{mu2022}, where a rhombohedral phase at low temperature has been determined, which corresponds to the high-symmetry phase determined here with a Ru--Cl--Ru bond angle of 93.73$^\circ$. This is almost 1$^\circ$ larger than the Ru--Cl--Ru bond angle found at 1.26\,GPa and room temperature, since $p$ indeed causes a reduction of Ru--Cl--Ru bond angle and brings it closer to the ideal value of 90$^\circ$.

From the above we conclude that biaxial stress can induce a structural transition from  $C2/m$ to the high-symmetry phase. On the one hand, this renders the high-symmetry phase with \textit{p}$\bar3$1\textit{m} layer symmetry accessible to a variety of experiments. On the other hand, this also means that care must be taken when mounting \rucl for low $T$ measurements. The strain-dependence may in fact be the reason why some previous studies reported the occurrence of a first-order structural phase transition during cooling~\cite{reschke2018,gass2020}.

It is remarkable that biaxial stress parallel to the \crc  affects the stacking in the perpendicular direction. The underlying physical mechanism certainly deserves further study. 
It will also be very important to determine the difference between the $C2/m$ and the high-symmetry phase in terms of magnetic couplings. Recent studies indeed report the relevance of  interlayer interactions for the magnetism of \rucl\,\cite{balz2021, Kaib:2021aa}. A very recent neutron scattering study could in fact identify the rhombohedral phase with the transition at T$_N$=7\,K (Ref.\,\onlinecite{Cao:2016a} and notes in Ref.\,\onlinecite{mu2022}). The latter corresponds to the high-symmetry polytype described above. It is therefore not unlikely that the transition at T$_N$=14\,K, which is observed in some samples, is related to the monoclinic $C2/m$-polytype.

To conclude, we showed that hydrostatic pressure $p$ as well as biaxial stress parallel to the $ab$-plane can drive the structure of \rucl closer towards the ideal Kiteav-geometry. An intricate interplay between the in-plane and out-of-plane structure in \rucl was observed, which not only modifies the magnetic interactions in the out-of-plane direction perpendicular to the \crc, but is likely to also affect the relative strength of the in-plane interactions and the magnetic ground state of this very promising Kitaev-candidate material. %
It will be extremely interesting to further explore this new aspect and its role for the QSL-behaviour in \rucl in future studies.

\section*{Acknowledgements}
We thank L. Janssen and M. Ruck for fruitful discussions. This research has been supported by the Deutsche Forschungsgemeinschaft through SFB 1143 (project-id 247310070), the W\"urzburg-Dresden Cluster of Excellence on Complexity and Topology in Quantum Matter–ct.qmat (EXC 2147, project-id 390858490) and SFB 1415 (project-id 417590517). We also gratefully acknowledge the support provided by the DRESDEN-concept alliance of research institutions and thank the ESRF for providing beamtime at ID27 and ID15B.
For their support during the synchrotron experiments we would like to thank M. L. Amig\'o, L. Lei\ss{}ner and T. Raman.

\section{Appendix}
 
\subsection{X-ray data analysis and structure refinement}

The data collected at ambient conditions for crystal 1 was processed using Bruker’s Apex3 software  (Ref.~\onlinecite{APEX}), the reflection intensities were integrated using SAINT (Ref.~\onlinecite{SAINT}) and multi-scan absorption correction was applied using SADABS (Ref.~\onlinecite{Krause2015}). The subsequent structure solution and weighted fullmatrix least-squares refinement on F$^2$ were done with SHELXT-2014/5 (Ref.~\onlinecite{Sheldrick2015a}) and SHELXL-2018/3 (Ref.~\onlinecite{Sheldrick2015}) as implemented in the WinGx 2018.3 program suite (Ref.~\onlinecite{Farrugia2012}). Key details of the data collection and the structural refinement are summarized in Table~\ref{tab:refine_amb}. The atomic positions and isotropic displacement parameter based on the single-crystal x-ray diffraction data at ambient conditions are listed in Table~\ref{tab:para_amb}.

To determine the averaged structure at 1.26\,GPa and ambient temperature of crystal 2, the data was integrated and corrected for Lorentz, polarization and background effects using the CrysAlisPro software suite (version 171.39.46)~\cite{CrysAlis}.
Reflections, which were saturated due to overexposure or an overlap with diamond peaks, were omitted from the integration process. Note that the separation of sharp and diffuse scattering requires no special treatment, as the Bragg peaks of family 1 and 2 are not affected by diffuse intensity. We collected 135 reflections, which were merged based on the crystal symmetry to 30 independent reflections with R$_{int}$ = 4.69\,\%. The averaged structure was then solved in the space group \textit{p}$\bar3$\textit{m}1 using the  SHELXT-2014/5 (Ref.~\onlinecite{Sheldrick2015a}) and SHELXL-2018/3 (Ref.~\onlinecite{Sheldrick2015}) as implemented in the WinGx 2018.3 program suite (Ref.~\onlinecite{Farrugia2012}).  The final refinement converged at R$_1$ (all data) = 2.65\,\% and wR$_{2}$ (all data) = 5.31\,\%. The parameters characterizing the data collection and the structural refinement are summarized in Table~\ref{tab:refine_hs}.

\begin{table*}
  \begin{ruledtabular}
    \caption{Details on data collection and structure refinement of \rucl as determined from single-crystal X-ray diffraction at ambient conditions.}
    \begin{tabular}{lclclc}
	\multicolumn{2}{l}{Crystal data} & \multicolumn{2}{l}{Data collection} & \multicolumn{2}{l}{Refinement}\\
	\hline
	\textit{Pressure}\,(GPa)  & 0 & \textit{Wavelength}\,(\AA) & 0.7107 & \textit{N$_{parameters}$} &22\\
	\textit{Temperature}\,(K)  & 295 & \textit{2$\theta_{max}$}\,($^\circ$) & 60.88 & \textit{R$_{1} > 4\sigma$}\,(\%) &1.62 \\
	\textit{Space group} & \textit{C}2/\textit{m} & $T_{min}$ & 0.6214 & \textit{R$_{1}$\,all}\,(\%) &1.62 \\
	\textit{a}\,(\AA) &  5.9875(6) & $T_{max}$ & 0.7461 & \textit{wR$_{2} > 4\sigma$}\,(\%) &3.86  \\
	\textit{b}\,(\AA) &  10.3529(3) & \textit{N$_{measured}$} & 2006 & \textit{wR$_{2}$ all}\,(\%) &3.86\\
	\textit{c}\,(\AA) &  6.0456(6) & \textit{N$_{observed}$} [\textit{I}$>$2$\sigma$(\textit{I})] & 555 & \textit{$\Delta\rho_{min}$} (e$\cdot$A$^{-3}$) &-0.773\\
	\textit{$\beta$}\,($^\circ$) &  108.777(9) & \textit{$\mu$}\,(mm$^{-1}$) & 6.397 & \textit{$\Delta\rho_{max}$} (e$\cdot$A$^{-3}$) &0.764 \\
	\textit{Z}  & 4 & \textit{R$_{int}$}  (\%) & 2.52 & \textit{G.O.F} &1.093 \\
	\textit{$\rho_{calc}$}\,(g$\cdot$cm$^{-3}$) & 3.883 & & &   \textit{weight w (a,b)} & 0.0195\\
	& & & & \textit{Extinction}&0.0007 \\
    \end{tabular}
    \label{tab:refine_amb}
  \end{ruledtabular}
\end{table*}

\begin{table*}[t!]
  \begin{ruledtabular}
   \caption{Details on data collection and structure refinement of \rucl as determined from single-crystal X-ray diffraction at 1.26\,GPa and ambient temperature.}
    \begin{tabular}{lclclc}
	\multicolumn{2}{l}{Crystal data} & \multicolumn{2}{l}{Data collection} & \multicolumn{2}{l}{Refinement} \\
	\hline
	\textit{Pressure}\,(GPa)  & 1.26 & \textit{Wavelength}\,(\AA) & 0.4113 & \textit{N$_{parameters}$} &6 \\
	\textit{Temperature}\,(K)  & 300 & \textit{2$\theta_{max}$}\,($^\circ$) & 37.5 & \textit{R$_{1} > 4\sigma$}\,(\%) &2.50 \\
	\textit{Space group} & \textit{P}$\bar3$\textit{m}1 & $T_{min}$ & 0.68 & 	\textit{R$_{1}$\,all}\,(\%) &2.65 \\
	\textit{a}\,(\AA) &  3.4080(4) & $T_{max}$ & 1.00 & \textit{wR$_{2} > 4\sigma$}\,(\%) &5.08 \\
	\textit{b}\,(\AA) &  3.4080(4) & \textit{N$_{measured}$} & 135 & \textit{wR$_{2}$ all}\,(\%) &5.31 \\
	\textit{c}\,(\AA) &  5.562(10) & \textit{N$_{observed}$} [\textit{I}$>$2$\sigma$(\textit{I})] & 30 & \textit{$\Delta\rho_{min}$} (e$\cdot$A$^{-3}$) &-0.458\\
	\textit{Z}  &  1 & \textit{$\mu$}\,(mm$^{-1}$) & 9.863 &  \textit{$\Delta\rho_{max}$} (e$\cdot$A$^{-3}$) &0.685  \\
	\textit{$\rho_{calc}$}\,(g$\cdot$cm$^{-3}$) & - & \textit{R$_{int}$}  (\%) & 4.69 &   \textit{G.O.F} &1.404   \\
	& & & &   \textit{weight w (a,b)} & 0.0191	\\
    \end{tabular}
    \label{tab:refine_hs}
  \end{ruledtabular}
\end{table*}

\begin{table}
  \begin{ruledtabular}
  \caption{Fractional atomic coordinates and equivalent isotropic displacement parameters (\AA$^{2}$) of a \rucl single crystal at ambient conditions.}
    \begin{tabular}{cccccc}
	Atom & Site & x & y & z & U$_{eq}$ \\
	\hline
	Ru & 4h & 0 & 0.16651(2) & 0.5 & 0.00964(9) \\
	Cl$_{1}$ & 4i & 0.22680(13) & 0 & 0.73488(12) & 0.01431(14) \\
	Cl$_{2}$ & 8j & 0.25058(10) & 0.17411(4) & 0.26761(9) & 0.01407(12) \\
    \end{tabular}
    \label{tab:para_amb}
  \end{ruledtabular}
\end{table}

\subsection{Extraction of the diffuse intensities}
The diffuse intensity profiles were extracted from the diffraction data collected with a MAR555 flat panel detector at ID15B of the ESRF. The diffraction images were transformed into reciprocal space and $\left|F^{2}_{0}\right|$  maps were reconstructed by applying Lorentz and polarization factors using the CrysAlisPro software package~\cite{CrysAlis}. 
For further data processing the Python packages numpy, matplotlib, and fabio were applied. The intensity profiles were estimated from the reconstructed layers for each pixel row along $l$ by adding the pixel values for all pixels lying within the peak region $h, k \pm\ 0.02$ and then subtracting a background intensity. The background intensity was as well determined line-by-line along $l$ by calculating the average intensity in regions $\Delta k$ = 0.03 immediately adjacent to the peak region.

\subsection{Modelling of the diffuse scattering}

\begin{figure}
 \includegraphics[width=0.9\columnwidth]{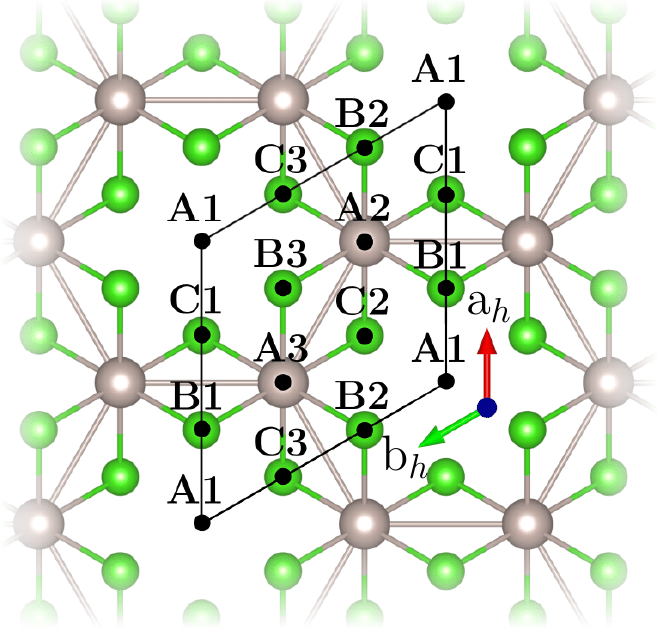}
 \caption{\textbf{Schematic illustration} Plane view of one Cl--Ru--Cl layer with in-plane basis vectors a$_h$ and b$_h$, and illustrated Ru honeycomb net. The A1, A2, ..., C3 notation is used to specify the nine different possibilities of stacking two adjacent layers on top of each other according to the position of the octahedral voids within a single layer.
}
\label{fig:Stacking}
\end{figure}

First, a single layer was built up by expanding the hexagonal unit cell to a 20x20x1 supercell. For the simulation 1000 of these layers were stacked along $c_h$. There are various possibilities to stack the individual layers, while preserving a hexagonal or cubic closed packing of the Cl-atoms. Altogether, there are 9 possibilities for the orientation of two adjacent layers that meet these requirements. The configuration of a single layer is specified by the position A1, A2, ..., C3 of the octahedral voids within the hexagonal cell, as illustrated in Fig.\,\ref{fig:Stacking}. In order to depict the stacking of the layers a stacking vector \textbf{T}$_S$ is implemented, which connects the octahedral voids in successive layers. The stacking vector \textbf{T}$_S$ with $S$=A1,A2,...C3 points from the octahrdal void A1 in the initial layer to an octahedral void $S$ in the successive layer. In the event of a hexagonal closed packing of the Cl-atoms the arrangement options for two consecutive layers is reduced to three. We define these three kinds of stacking as $eclipsed$ ($e$), $forward$ ($f$) and $backward$ ($b$), that correspond to the stacking vectors \textbf{T}$_{A1}$ = [0,0,1], \textbf{T}$_{A2}$ = [2/3,1/3,1] and \textbf{T}$_{A3}$ = [1/3,2/3,1] respectively. However, our analysis showed that the measured XRD-data is very well reproduced by the exclusive use of $f$ and $b$ stacks. In order to probe the short-range order in the pressurized \rucl samples, the sequence of $f$ and $b$ stacks in our simulation is generated by a first-order Markov process. To resemble the experimental found stacking disorder two independent transition probabilities $p_{ff}$ and $p_{bb}$ must be defined. Here, $p_{ff}$ corresponds to the probability that a layer be $f$ stacked on a preceding $f$ stack, and accordingly $p_{bb}$ is the probability of continuing in a $backward$ stacking sequence. Therefore the probabilities for the presence of a stacking fault are $p_{fb}$=1-$p_{ff}$ and $p_{bf}$=1-$p_{bb}$, respectively. The process is characterized in form of a right stochastic matrix \textbf{P} containing the transition probabilities.

\begin{align*}
\mathbf{P} = 
\begin{pmatrix*}[r]
   p_{ff} & p_{fb}\\[3pt]
   p_{bf} & p_{bb}\\[3pt]
\end{pmatrix*}
\end{align*}

Using this approach DISCUS creates a list of layer positions. The scattering intensity of the layered model crystal is finally calculated as the product of the individual Fourier transform of the layer positions and the single layer. All in all we created 676 different disorder models by adjusting the transition probabilities $p_{ff}$ and $p_{bb}$ in steps of 0.04. For each model we calculated the intensity profiles along 12$l$, 1-3$l$ and 1-4$l$ over the range -2 $\leq$ $l$ $\leq$ +2. Obviously the model crystal contain far fewer layers than do real crystals and consequently the number of stacking fault events is small. To reduce the statistical noise and thus produce a smooth intensity distribution along $l$ the line profiles for each set of transition probabilities $p_{ff}$ and $p_{bb}$ were simulated 200 times and merged. To obtain quantitative agreement between simulated and measured XRD-data, the calculated diffuse profiles were adjusted by a scale factor to the observed intensity profiles.

The examination of the simulated intensity profiles revealed that the stacking probabilities and the $x$ coordinate of the Cl-atom have significantly different influence on the distribution of the intensity along $l$, and can therefore be determined separately.The probabilities of finding a $t$ or $f$ stack determine the shape and position of the peak maxima.
Even small variations in the probability values induce discernible effects on the simulated intensity profiles. We estimate the uncertainties of the optimized values $p_{ff}$ = 0.60 and $p_{bb}$ = 0.72 to be $\pm$ 0.02. In contrast, the relative intensities with a period $l=n+1$ along the streaks are mainly dependent on the $x$ coordinate of the Cl-atom, which is particularly evident for the intensity maxima located at $l=n+1/3$ and $l=n+2/3$. The simulation was carried out for several $Cl$-positions $\Delta x$ displaced in 3\,pm steps parallel to $a_h$ from the average position $x_{Cl}$ = 1/3 ($\Delta x$ = 0\,pm) deduced from the sharp reflections. The results are shown in Fig.~\ref{fig:hexagonal} (a)-(c) by solid lines. Note that only the position of the Cl-atom is changed for the simulated profiles shown in Fig.~\ref{fig:hexagonal} (a)-(c), while the stacking fault probabilities remain unaltered.


%

\end{document}